\documentclass[journal,comsoc]{IEEEtran}
\usepackage{siunitx} 
\IEEEoverridecommandlockouts

\usepackage{cite}
\usepackage{physics}       
\usepackage{amsmath,amssymb,amsfonts}
\usepackage{algorithmic}
\usepackage{graphicx}
\usepackage{textcomp}
\usepackage{xcolor}
\usepackage{cutwin} 
\usepackage[font=footnotesize,labelfont=bf]{caption}  
\usepackage[T1]{fontenc}
\usepackage[colorlinks=true, linkcolor=black, citecolor=black, urlcolor=black]{hyperref}

\usepackage[capitalise]{cleveref}
\usepackage{soul}

\begin{document}

\title{Entanglement Purification With Finite Latency Classical Communication in Quantum Networks} 

\author{Vivek Vasan\IEEEauthorrefmark{2}, Alexander Nico-Katz\IEEEauthorrefmark{3}$^{,}$\IEEEauthorrefmark{1}, Boulat A. Bash\IEEEauthorrefmark{5}, Daniel C. Kilper\IEEEauthorrefmark{4}, and Marco Ruffini\IEEEauthorrefmark{2}
	\vspace{0.5em}
	\\ \IEEEauthorrefmark{2}School of Computer Science and Statistics, CONNECT Centre, Trinity College Dublin, Dublin, Ireland\\\IEEEauthorrefmark{3}School of Physics, Trinity College Dublin, Dublin, Ireland\\\IEEEauthorrefmark{4}Department of Electronic and Electrical Engineering, CONNECT Centre, Trinity College Dublin, Dublin, Ireland\\\IEEEauthorrefmark{5}Department of Electrical and Computer Engineering, University of Arizona, Tucson, AZ, USA\\\IEEEauthorrefmark{1}Trinity Quantum Alliance, Unit 16, Trinity Technology and Enterprise Centre, Pearse Street, Dublin 2, Ireland \\
		\textit{Corresponding author email}: vasanv@tcd.ie
}

\maketitle

\begin{abstract}
Quantum networks rely on high fidelity entangled pairs distributed to nodes, but maintaining their fidelity is challenged by environmental decoherence during storage. Entanglement purification is used to restore fidelity, but the idle periods imposed by the associated classical communication delays counteract this goal by exposing the states to further decoherence.
In this work, we analyze the practical viability of entanglement purification protocols (BBPSSW, DEJMPS), under non-instantaneous classical coordination over Internet protocol (IP) communications networks.
We present a comprehensive performance evaluation of these protocols in various network conditions for a range of quantum memory technologies. We employ a microscopic Lindblad treatment of the underlying quantum dynamics, and use current-generation metropolitan IP network latency statistics and parameters drawn from quantum memory testbeds. In doing so we identify the regions in which entanglement purification succeeds and fails, delineated by break-even iso-fidelity contours in the phase space. We then determine the total number of entangled pairs required to complete a multi-round purification protocol, and the steady-state throughput of entangled pairs with purified fidelities that exceed application-specific thresholds. This provides latency budgets, memory quality targets, and resource-overhead estimates for deploying purification on current and near-future networks.



\end{abstract}

\section{Introduction}
\label{sec:intro}

Quantum networks represent a paradigm shift in the information technology space and are poised to revolutionize the way information is transmitted and secured. By harnessing quantum mechanical principles, they unlock capabilities with no classical analogue, enabling applications such as quantum key distribution (QKD), distributed quantum computing (DQC), and quantum sensing \cite{van2014quantum}. The foundational resource for these networks is high-fidelity quantum entanglement distributed between distant nodes in a network. In practice these entangled states are fragile and degrade (losing their characteristic quantum features) due to interactions with the environment: a process known as decoherence. The primary operational challenge, therefore, is to counteract this inevitable degradation. Entanglement purification protocols (EPPs) are the canonical solution, offering a method to distill a small number of high-fidelity entangled pairs from a larger ensemble of low-fidelity pairs \cite{dur2007entanglement}. 

However, a critical and often idealized aspect of these protocols is their reliance on classical communication to coordinate local quantum operations. This coordination occurs over auxiliary channels, which play an essential supporting role to the primary quantum channel. In any practical implementation, these auxiliary channels are realized using conventional networks, introducing finite and variable latency. This creates a fundamental trade-off: the very protocol designed to increase fidelity forces quantum states to be stored in imperfect memories for the duration of the classical communication, subjecting them to non-trivial decoherence. From a systems engineering perspective, it is therefore critical to consider how this latency impacts the network's quality of service. This ``race against time'' between fidelity gain due to purification and latency-induced fidelity loss is the central problem that this work addresses. In this context, three natural questions emerge: (i) under the constraints of realistic network latency and imperfect quantum hardware, what are the precise operational boundaries that determine whether EPPs provide a net fidelity gain, net fidelity loss, or break-even? (ii) How do these boundaries shift as a function of network contexts and choice of purification protocol? (iii) What are the ultimate operational limits on the achievable rate of high-fidelity entangled pairs that can be delivered to an application and what are the associated resource costs in terms of Bell pair consumption? 

To address these questions, we present a detailed systems-level analysis of two seminal recurrence-based EPPs, the Bennett-Brassard-Popescu-Schumacher-Smolin-Wootters (BBPSSW \cite{bennett1996purification}) protocol and the Deutsch-Ekert-Jozsa-Macchiavello-Popescu-Sanpera (DEJMPS \cite{deutsch1996quantum})) protocol described in detail in \cref{sec:distillation}. Crucially, whilst previous analyses idealize classical communication as either instantaneous or contained within a small enough decoherence window as to be negligible when compared to the overhead from, e.g., swapping and pair generation, real world networks exhibit significant and variable signaling delays which cannot be neglected. We address this in our study by analyzing network performance at a series of fixed classical latencies.
These are sampled from an empirical distribution derived from a real-world metropolitan Internet protocol (IP) network, which ensures they are representative of a wide range of network conditions (see \cref{sec:model}). Additionally, we explicitly model the continuous-time dynamics of qubits stored in quantum memories using a Lindblad master equation: a rigorous physical model which we parameterize using the experimentally characterized $T_1$ and $T_2$ times of specific memory technologies. Finally, we focus on purification between end nodes of a network, contrasting with extant works (inter alia, Refs.~\cite{brand2020efficient, li2021efficient, victora2023entanglement, zang2025entanglement}) which probe fidelity loss during idling in repeater chains. Our work, taken in its totality, evaluates established purification protocols in detailed treatments of current-generation or near-future networking contexts. This synthesis of a microscopic decoherence model with real-world network data allows for the capture of the full stochastic impact of the classical control plane on the performance of quantum protocols, providing a far more realistic assessment of their practical viability than has been previously available.

Our findings in \cref{sec:results} provide a clear and quantitative picture of the operational regimes for entanglement purification in practical networks. First, we identify sharp ``break-even'' boundaries in the parameter space of latency and resource overhead (measured as expected number of noisy bell pairs consumed), providing a clear map of the regimes where purification is beneficial versus where it fails and degrades fidelity below the initial state. Second, we quantify the consistent and significant performance advantage of the DEJMPS protocol over BBPSSW under these realistic conditions. For any single round of purification, DEJMPS is known to yield a marginally higher fidelity gain and success probability than BBPSSW.  However, this marginal advantage results in DEJMPS delivering steady-state end-to-end distillable entangled pair rates that are often orders of magnitude greater, particularly at low-to-moderate latencies.  Finally in \cref{sec:conclusion}, we translate our results into concrete design rules for network engineers, including practical latency budgets, minimum required memory coherence times, and explicit resource overhead estimates needed to meet the fidelity thresholds for quantum network applications considered. Ultimately, these constitute an essential guideline for the engineering and deployment of practical, large-scale quantum networks, transforming abstract protocol analysis into actionable system design parameters.



\section{Quantum Networking Model and Motivation}
\label{sec:model}

\begin{figure*}[!ht]
    \centering
    \includegraphics[width=\textwidth]{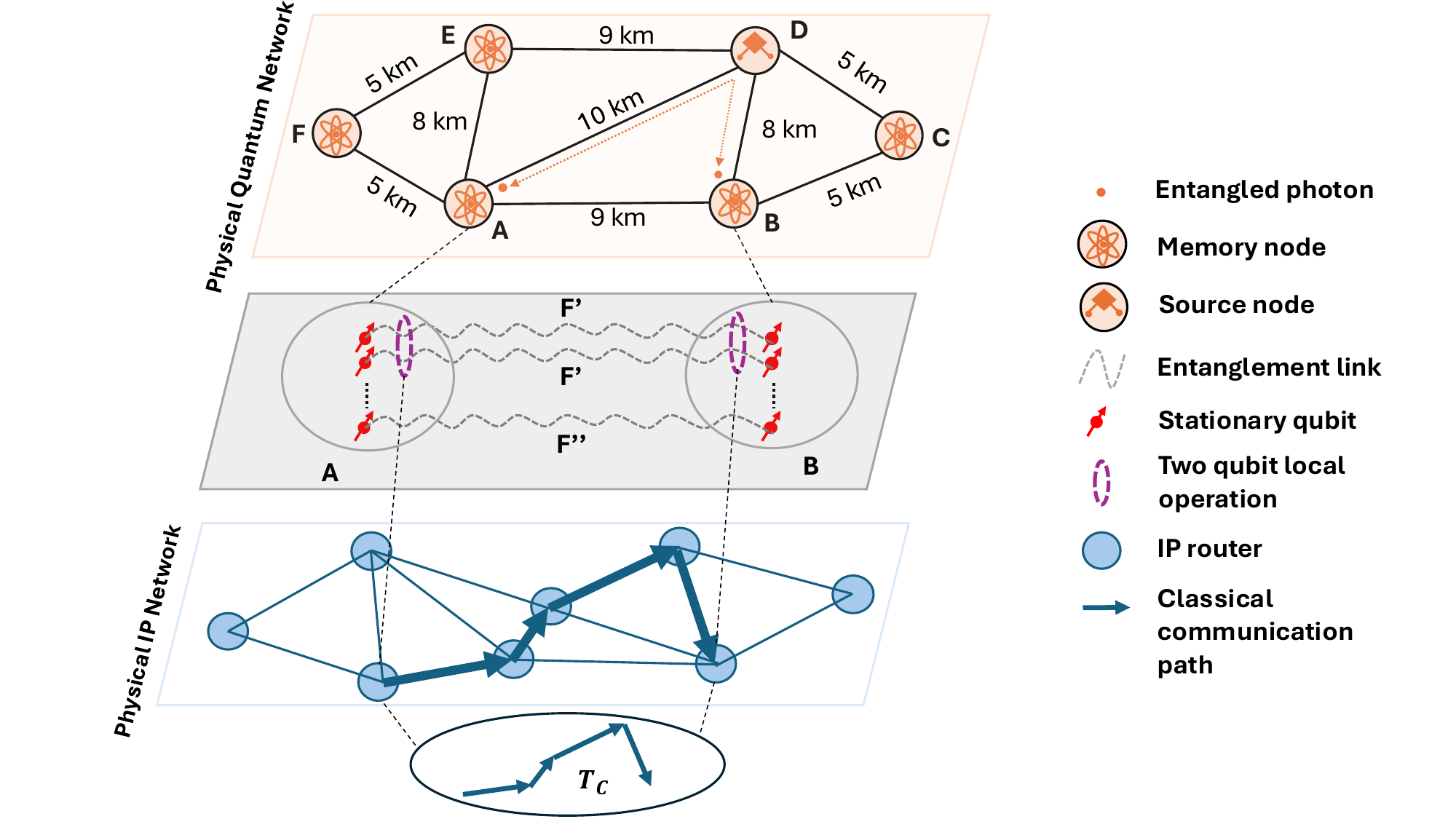}
    \caption{Entangled photon pairs are distributed and mapped into quantum memories at nodes A and B, creating multiple bipartite links of certain fidelity. Purification proceeds by performing local two‐qubit operations on matching‐fidelity pairs and exchanging the resulting measurement outcomes over the classical IP network. If successful, purification  boosts the fidelity ($F'' > F'$) as denoted in the figure. The latency of this classical signaling is denoted by $T_C$. }
    \label{fig:schematic}
\end{figure*}
\subsection{Classical Control Plane for Quantum Network}

We consider a point-to-point entanglement distribution link between two quantum nodes (which we will call node A and node B) as shown in \cref{fig:schematic}. The link is realised with optical photons as carriers of entanglement over long distances due to its robustness to environmental noise. At each node, the photonic qubits can be readily entangled with  or mapped onto atomic or solid-state quantum memories, enabling applications to consume the distributed entanglement. These nodes have interfaces  connecting two types of channels: (1) a quantum channel (i.e., optical fiber) over which entangled photons are transmitted from a source to both nodes, and (2) a classical channel for the control and coordination  between the nodes.  In a practical deployment, the latter channel is realized using conventional networking technology, such as a metropolitan-scale IP network. 

The classical channels enable the network control plane. In classical networking, the control plane is responsible for the decision-making and management logic of the network, such as determining routes, managing connections, and signaling, while the data plane is responsible for the actual forwarding of data packets. Analogously, quantum network uses its classical control plane to coordinate tasks such as negotiating connection setups, heralding successful entanglement outcomes, initiating entanglement swapping, and managing resource allocation (notably quantum-memory usage). In our design, the classical channel is an IP network path, which is physically separate from the quantum fiber (for instance, running over Internet links). Sending classical signals over the same fiber as quantum photons is typically avoided because they induce noise (e.g., Raman scattering) that can disrupt the quantum channel, so the control traffic is offloaded to a legacy IP network. While practical, this introduces non-negligible and variable latency. 

\begin{figure}[!ht]
    \centering
    \includegraphics[width=\linewidth]{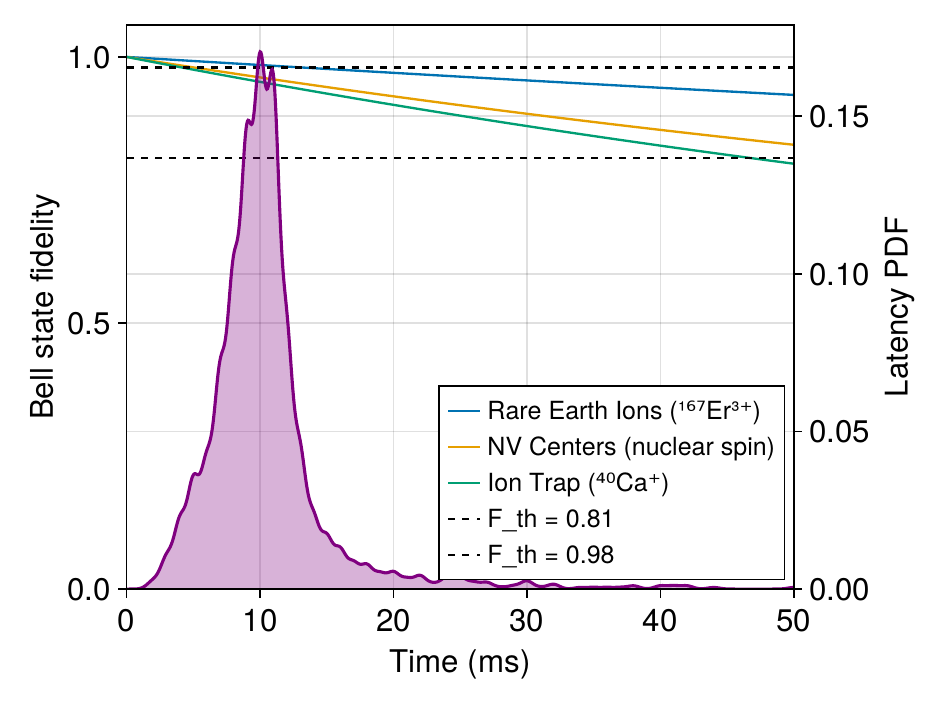}
    \caption{\textbf{Fidelity degradation due to control plane latency in realistic quantum memories}. State fidelity \(\mathcal{F}(T_C)\) vs.\ classical latency for various \(T_1,T_2\). The $0.81$ QKD threshold and $0.98$ DQC threshold are shown, and the shaded purple region is the IP network latency pdf for the Dublin‐metro area.}
    \label{fig:fidelity-fixed_latency}
\end{figure}

The groundwork for the present analysis was established in our prior conference paper \cite{vasan2024control}, which proposed a control plane protocol running on a legacy IP network to verify entanglement pairing (i.e., recognize and acknowledge that two nodes share a quantum link) before they can be consumed for application. That study quantified the impact of control-plane delays on end-to-end fidelity by modeling the classical channel with empirical latency data from a metropolitan network (Greater Dublin Area, Q1 2024), provided by Ookla's open data initiative \cite{ookla2024speedtest}. Throughout our simulations, we assume symmetric one-way signaling times ($\tau_{A\to B} = \tau_{B \to A} = T_C$). Although individual packet delays may differ in practice due to router-specific queuing behaviors, we conservatively use the higher latency value from the two directions. The probability density function of this dataset along with the fidelity degradation curves for the three representative quantum memory platforms listed in \cref{tab:tab1} is shown in \cref{fig:fidelity-fixed_latency}. The two horizontal dotted lines at $0.81$, $0.98$ fidelity denote the thresholds required for the quantum key distribution (QKD) \cite{wengerowsky2018entanglement} and distributed quantum computing (DQC) \cite{jacinto2025network}, respectively. 


\begin{table}[htbp]
\caption{Experimental parameters for different quantum memory technologies\vspace{-1em}}
\label{tab:tab1}
\begin{center}
\begin{tabular}{|c|c|c|}
\hline
\textbf{Quantum Memory Technology} & \multicolumn{2}{|c|}{\textbf{Parameters}} \\
\cline{2-3}
& \textbf{\textit{T\textsubscript{1} (s)}} & \textbf{\textit{T\textsubscript{2} (s)}} \\
\hline
Ion Trap (\(^{171}\text{Yb}^+\)) & 12000\textsuperscript{b} & 4200\textsuperscript{b} \\
\hline
Rare Earth Ions (\(^{167}\text{Er}^{3+}:Y_2SiO_5\)) & 600\textsuperscript{c} & 1.3\textsuperscript{c} \\
\hline
Ion Trap (\(^{40}\text{Ca}^+\)) & 1.14\textsuperscript{a} & 0.5\textsuperscript{a} \\
\hline
NV Centers in Diamond (Nuclear Spin) & 200\textsuperscript{d} & 0.5\textsuperscript{d} \\
\hline
Superconductor Cavity & 0.0256\textsuperscript{e} & 0.034\textsuperscript{e} \\
\hline
Superconductor Cavity & 0.0012\textsuperscript{f} & 0.00072\textsuperscript{f} \\
\hline
\multicolumn{3}{l}{\textsuperscript{a}\cite{kreuter2005experimental}, \textsuperscript{b}\cite{wang2021single}, \textsuperscript{c}\cite{rancic2018coherence}, \textsuperscript{d}\cite{maurer2012room}, \textsuperscript{e}\cite{milul2023superconducting}, \textsuperscript{f}\cite{reagor2016quantum}}
\end{tabular}
\end{center}
\vspace{-2em}
\end{table}

In light of the fidelity limitations inherent to realistic quantum networks, entanglement purification protocols are essential to extend the operational regime and satisfy application specific requirements. The current work invesitgates the extent to which entanglement purification can counteract the fidelity loss experienced by entangled pairs in the network. 
We identify three primary stages where this entanglement degradation occurs: (i) in flight fiber noise, (ii) errors introduced during the quantum transduction process (i.e., mapping the photonic state to the quantum memory), and (iii) decoherence accumulated as a function of storage time while awaiting classical coordination signals. Our prior work \cite{vasan2024control} specifically quantified this third component, demonstrating that latency induced storage decoherence is a critical factor. In this study, we encapsulate the combined effect of all three degradation sources into a single parameter, the initial fidelity, $F_{0}$. This value represents the realistic, imperfect state of an entangled pair at the moment a purification protocol would begin, setting the stage for our analysis.
In the following sections, we model continuous time decoherence via a Lindblad master equation (see \cref{sec:lindblad}) and introduce and evaluate the BBPSSW and DEJMPS purification protocols (see \cref{sec:dist-protocols}) to quantify their performance in terms of achievable fidelity, throughput and the associated resource overhead as a function of latency under realistic metro network conditions.

\subsection{Lindblad Treatment of Continuous-Time Decoherence}
\label{sec:lindblad}
Here we construct a single-step Lindblad treatment of the Markovian dynamics of qubits stored in memory, i.e., a setting wherein information lost to the environment cannot be restored. Following our previous work in ~\cite{vasan2024control}, we consider standard quantum memory registers which exhibit amplitude damping and phase errors governed by their respective $T_1$ and $T_2$ relaxation times (see \cref{tab:tab1}). Furthermore we assume these errors to be identical at all nodes although an extension to node-dependent decoherence times is relatively straightforward. We first construct explicit jump operators to characterize the relevant dynamical processes
\begin{equation}
    \hat{L}_{1,A(B)} = \sqrt{\gamma_1} \hat{\sigma}^-_{A(B)}, \quad \hat{L}_{2,A(B)} = \sqrt{\gamma_2} \hat{\sigma}^z_{A(B)},
\end{equation}
local to nodes $A$ and $B$, with decay rates given by $\gamma_{j} = 1/T_j$ \cite{singh2020using}, and where $\hat{\sigma}^z_{A(B)}$ and $\hat{\sigma}^-_{A(B)}$ are the standard Pauli $z$-operator and lowering operators, acting on nodes $A(B)$ respectively. We can then track a state's dynamics by integrating the standard Lindblad equation \cite{breuer2002theory}:
\begin{equation}\label{eq:lindblad}
    \partial_t \hat{\rho}(t) = -i[\hat{H}, \hat{\rho}(t)] + \sum_{j,\alpha} \mathcal{D}\left[\hat{L}_{j,\alpha}, \hat{\rho}(t)\right],
\end{equation}
where $j\in\{1,2\}$, $\alpha\in\{A, B\}$, the commutator describes the unitary component of the state's evolution, and the dissipators account for the non-unitary dynamics induced by the jump operators and are given by
\begin{equation}
    \mathcal{D}\left[\hat{L}_{j,\alpha}, \hat{\rho}(t)\right] = \hat{L}_{j,\alpha} \hat{\rho}(t) \hat{L}_{j,\alpha}^\dagger - \frac{1}{2} \{ \hat{L}_{j,\alpha}^\dagger \hat{L}_{j,\alpha}, \hat{\rho}(t) \}.
\end{equation}
The unitary component is governed by the trivial Hamiltonian $\hat{H} = \mathbb{I}_A\otimes\mathbb{I}_B$, corresponding to an ideal memory realizing trivial evolution, i.e., idling. Integrating the Lindblad equation of \cref{eq:lindblad} can be numerically intensive though this is mitigated in our work by the fact that relevant dynamical timescales (the $~10$ms latency times) are orders of magnitude smaller than the relevant $T_1$ and $T_2$ times, enabling the use of a single-step fourth order Runge-Kutta integration method (see \cref{sec:convergence} for discussion and convergence checks). The other prevailing method of Lindblad master equation integration, the trajectory unravelling method, is irrelevant here due to the aforementioned relative timescales and the small system sizes we consider, but we mention it for completeness and refer the interested reader to Ref.~\cite{barchielli2009quantum, breuer2002theory}.

\section{Description and modeling of the entanglement purification protocols}
\label{sec:distillation}

In this section, we present the theoretical framework underpinning the use of purification protocols to improve (and then maintain) fidelity of entangled pairs in the context of the long-range quantum networks introduced in \cref{sec:model}. 

\subsection{Verified memory states}
\label{sec:werner-and-fidelity}

As outlined in \cref{sec:model}, and examined in detail in Ref.~\cite{vasan2024control}, an entangled pair may have significantly degraded by the time each flying qubit has arrived, been converted to stationary memory, and been subsequently verified and paired via a classical exchange of information. This degradation can be captured by an ansatz state of Werner form:
\begin{equation}
    \hat{\rho}_W(F) = F\ket{\Psi^-}\bra{\Psi^-} + \frac{1-F}{3}\left(\mathbb{I}-\ket{\Psi^-}\bra{\Psi^-}\right).
    \label{eq:werner}
\end{equation}
where $|\Psi^-\rangle$ is the Bell singlet state, and where we have (implicitly) introduced the Bell states:
\begin{subequations}
\begin{align}
    \ket{\Psi^\pm} &= \frac{1}{\sqrt{2}}\left(\ket{0}_A \ket{1}_B \pm \ket{1}_A \ket{0}_B\right), \\
    \ket{\Phi^\pm} &= \frac{1}{\sqrt{2}}\left(\ket{0}_A \ket{0}_B \pm \ket{1}_A \ket{1}_B\right).
\end{align}
\end{subequations}
For notational clarity we often suppress the subscripts denoting which qubits belong to the spaces $A$ and $B$ (corresponding to local memories on nodes $A$ and $B$ respectively) except where these subscripts are critical to an understanding of the purification protocols (e.g., in the enactment of bilateral CNOT operations).

The Werner form of \cref{eq:werner} conceptually reflects a state which has a definite fidelity $F$ with the (target) singlet state, with the remaining amplitudes in a uniform mixture of states orthogonal to said singlet. Explicitly, for the fidelity function
\begin{equation}
    \mathcal{F}(\hat\rho, \hat\sigma) = \left(\Tr\sqrt{\hat\rho^{1/2}\hat\sigma\hat\rho^{1/2}}\right)^2,
\end{equation}
the parameter $F$ of a quantum state $\hat\rho$ is extracted as $F = \mathcal{F}(\hat\rho, \ket{\Psi^-}\bra{\Psi^-})$. Whilst not formally a metric, the fidelity function $\mathcal{F}$ is bounded between zero and unity, is symmetric in its arguments, and can be interpreted as the probability that one its arguments can pass for the other in the course of some protocol or measurement. This makes it useful for quantifying the `closeness' of two quantum states. Hence, we employ $F$ as a figure of merit for quantifying the quality of distributed Bell pairs across nodes in a network. 

Throughout this work, wherever we generate random initial states, we do so by generating a random $4 \times 4$ trace-unity real diagonal matrix $\hat \rho_D$ and subject it to a random rotation drawn from the $U(4)$ Haar ensemble $U_H^{[4]} \sim \mathrm{Haar}(4)$, $\hat \rho_R = U_H^{[4]} \rho_D U_H^{[4]\dagger}$. We then decompose the random state $\hat\rho_R$ into components parallel $\hat\rho_P$ and orthogonal $\hat\rho_O$ to the Bell singlet and construct the initial state as $\hat\rho = F \hat\rho_P + (1-F)\hat\rho_O$. This construction generates a random state similar to $\hat \rho_W(F)$ but with an uneven mixture of the non-singlet components. This is more realistic to noise models which do not preserve the Werner form, and which highlights the importance of twirling in the BBPSSW protocol (see \cref{sec:bennett-protocol}).

\subsection{Twirling Operations}
\label{sec:twirling}

The Werner state of \cref{eq:werner} does not just reflect a desirable ansatz state, but indeed can be explicitly brought about by subjecting an arbitrary state $\hat\rho$ to random or deterministic local rotations, this process is called `twirling' in literature. We introduce, and briefly discuss advantages and disadvantages of two important twirling operations here: (i) the deterministic protocol, and (ii) the Haar random protocol. In both cases we consider an arbitrary quantum state $\hat\rho$ with fidelity $F = \mathcal{F}(\hat\rho, |\Psi^-\rangle\langle\Psi^-|)$ with the Bell singlet.

First we consider (i) the deterministic twirling protocol enacted by the Kraus map
\begin{equation}\label{eq:kraus-twirl}
    \hat\rho_W(F) = \frac{1}{12}\sum_{i=1}^3\hat T_j\left(\sum_{j=1}^4 \hat T_j \hat T_j \hat\rho \hat T_j^\dagger\hat T_j^\dagger\right)\hat T_j^\dagger,
\end{equation}
where the Kraus operators $\hat T_j = \hat u_j \otimes \hat u_j$ are given by the $\hat u_j$,
\begin{align}
    \hat u_1 &= (\mathbb{I}+i\hat\sigma_x)/\sqrt{2}, \label{eq:u1-kraus} \\
    \hat u_2 &= (\mathbb{I}-i\hat \sigma_y)/\sqrt{2}, \\
    \hat u_3 &= i \ket{0}\bra{0}+\ket{1}\bra{1} ,\\
    \hat u_4 &= \mathbb{I}
\end{align}
and where the $\hat \sigma_\alpha$ are the standard $\mathrm{SU}(2)$ Pauli matrices. The map of \cref{eq:kraus-twirl} leaves the singlet component of $\hat\rho$ unchanged and sends the other components of the state to a uniform mixture of states orthogonal to the singlet, thus bringing about the Bell-Werner form of \cref{eq:werner}. The major advantages of this twirling protocol are that it is deterministic, and that it can yield more distilled entanglement in the context of certain protocols (see \cref{sec:bennett-protocol}). The major drawback is that the map of \cref{eq:kraus-twirl} is difficult to implement in practice.

Secondly we consider (ii) the Haar random twirling protocol which simply corresponds to drawing a matrix $\hat{U}_H \sim \mathrm{Haar}(2)$ drawn from the Haar measure over the group $\mathrm{U}(2)$ of $2\times2$ unitary matrices. The twirling protocol is then enacted by simply performing identical rotations on both parties $A$ and $B$ of the state $\hat\rho$ as follows:
\begin{equation}\label{eq:haar-twirl}
    \hat\rho_W(F) = \left(\hat U_H \otimes \hat U_H \right) \hat\rho  \left(\hat U_H \otimes \hat U_H \right)^\dagger,
\end{equation}
which leaves the singlet component of $\hat\rho$ unchanged, but mixes the triplet components. The major advantages of this twirling protocol is that it is considerably simpler to implement -- requiring only a single random rotation of each qubit. The major drawbacks are that this protocol is random, that parties at both $A$ and $B$ must prepare \textit{identical} random unitary rotations (or high-fidelity $k$-designs thereof), and that these random rotations may yield less distilled entanglement in the context of certain protocols (see \cref{sec:bennett-protocol} and discussion of twirling in Ref.~\cite{bennett1996mixed} and footnote 33 therein).

\subsection{Purification Protocols}
\label{sec:dist-protocols}
Here we introduce both the Bennett-Brassard-Popescu-Schumacher-Smolin-Wootters (BBPSSW) and the Deutsch–Ekert–Josza–Macchiavello–Popescu–Sanpera (DEJMPS) entanglement purification protocols for pairs of entangled qubits. While more sophisticated protocols exist, BBPSSW and DEJMPS are widely used as toy protocols in literature, and serve as ideal testbeds in which to map out the competition between entanglement purification and environmengtal decoherence in quantum networks. 

We first introduce the quantum-mechanical spaces relevant to both: namely two spaces $\mathcal{D}(\mathcal{H}_{AB})$ and $\mathcal{D}(\mathcal{H}_{A^\prime B^\prime})$ of physical density operators on the `source' $\mathcal{H}_{A B} = \mathcal{H}_{A}\otimes\mathcal{H}_{B}$ and `target' Hilbert spaces $\mathcal{H}_{A^\prime B^\prime} = \mathcal{H}_{A^\prime}\otimes\mathcal{H}_{B^\prime}$, respectively. Conceptually, $\mathcal{H}_A$ and $\mathcal{H}_{A^\prime}$ correspond to memory qubit spaces local to node $A$, and $\mathcal{H}_B$ and $\mathcal{H}_{B^\prime}$ correspond to memory qubit spaces local to node $B$. 


Both protocols consist of an initial series of operations (which are deterministic except where random twirling is used) performed locally by nodes $A$ and $B$ on their respective source and target qubits, followed by local measurements on the target qubits, an exchange of $Z$-basis measurement outcomes via classical channel between nodes $A$ and $B$, and then some post-processing upon protocol success (in both BBPSSW and DEJMPS, success corresponds to matching measurement outcomes). We close this section with a discussion of the differences between the two protocols, and some supporting numerical analyses to lay the groundwork for the more comprehensive results of \cref{sec:results}.

\subsubsection{BBPSSW Protocol}
\label{sec:bennett-protocol}

The BBPSSW entanglement purification protocol, as initially proposed by its namesakes in Ref.~\cite{bennett1996purification}, maps two identical states of Werner form $\hat\rho_W(F)$ (or, equivalently, two unknown states drawn from an ensemble with Werner form mean) to a new state $\hat\rho_W(F^\prime)$ with a certain probability $p(F)$ of success. If the source does not produce Werner form states, we assume without loss of generality that both nodes conspire to use either deterministic (see \cref{eq:kraus-twirl}) or random (see \cref{eq:haar-twirl}) twirling to bring both entangled pairs into the Werner form of \cref{eq:werner}. In order to achieve a probabilistic fidelity gain the initial fidelity must exceed $F \geq 1/2$ (see \cref{sec:protocol-compare}. The BBPSSW protocol now proceeds as follows:
\begin{enumerate}
    \item Node $A$ performs a unilateral $Y$-rotation of \textit{both} of its local memory qubits,
    \begin{equation}
        \hat\rho_{\widetilde{W}}(F) = (\hat\sigma_y\otimes\mathbb{I})\hat\rho_W(F)(\hat\sigma_y\otimes\mathbb{I}).
    \end{equation}
    The effect of such a rotation is to recast an object in the Werner form of \cref{eq:werner} with a fidelity $F$ with the singlet state, to an equivalent Werner form (denoted by a tilde) with a fidelity $F$ with the triplet state $|\Phi^+\rangle$.
    \item Nodes $A$ and $B$ both perform controlled-NOT (CNOT) operation between their respective source and target qubits, the combined process being the bilateral exclusive OR (BXOR): $\widehat{\mathrm{BXOR}} = \mathrm{CNOT}_{A \to A'} \mathrm{CNOT}_{B \to B'}$. The combined state at this stage is given by: 
    \begin{equation}
        \hat\rho_\mathrm{MX}(F) = \widehat{\mathrm{BXOR}}~\hat\rho_{\widetilde{W}}(F) \otimes \hat\rho_{\widetilde{W}}(F)~\widehat{\mathrm{BXOR}}^\dagger,
    \end{equation}
    where the subscript MX denotes the fact that this is the final deterministically induced state of the system before the probabilistic measure-and-exchange (MX) step.
    \item Nodes $A$ and $B$ both perform $Z$-basis measurements on their respective target qubits and exchange the measurement outcomes via a classical channel. With probability $p(F)$ the measurement outcomes match and the protocol continues, with probability $1-p(F)$ they do not match and both nodes discard their respective source qubits. The POVM elements corresponding to successful protocol outcomes are
    \begin{align}
        \hat P_0 &= \mathbb{I}_A\otimes\mathbb{I}_B\otimes\ket{0}\bra{0}_{A'}\otimes \ket{0}\bra{0}_{B'} ,\\ 
        \hat P_1 &= \mathbb{I}_A\otimes\mathbb{I}_B\otimes\ket{1}\bra{1}_{A'}\otimes \ket{1}\bra{1}_{B'},
    \end{align}
    such that the protocol succeeds with total probability:
    \begin{equation}
    p(F) = \sum_{j\in\{0,1\}} \Tr\left(\hat{P}_j \hat{\rho}_\mathrm{MX}(F) \hat{P}_j^\dagger \right),
    \label{eq:success-probability}
\end{equation}
    with final state (after a successful protocol) given by:
    \begin{equation}
        \hat\rho_\mathrm{succ}(F) = \frac{1}{p(F)}\sum_{j\in\{0,1\}} \hat P_j \hat\rho_\mathrm{MX}(F) \hat P _j^\dagger.
    \end{equation}
    \item If the measurement outcomes match (with probability $p(F)$), both parties keep their respective source qubits and reverse the initial $Y$-rotation
    \begin{equation}\label{eq:inverse-y-rot}
        \hat\rho(F') = (\hat\sigma_y\otimes\mathbb{I})\hat\rho_\mathrm{succ}(F)(\hat\sigma_y\otimes\mathbb{I}),
    \end{equation}
    to retrieve a state $\hat\rho(F')$ with increased fidelity $F' > F$ with the Bell singlet. Optionally, this output state can be twirled to bring it back to Werner form.
\end{enumerate}
For deterministic twirling, the value of the increased fidelity $F'$ can be calculated analytically (see Ref.~\cite{bennett1996purification} for details) as follows: 
\begin{equation}\label{eq:analytic-bennett}
    F' = \frac{F^2 - (1-F)^{2/9}}{F^2 + 2 F(1-F)/3 + 5(1-F)^{2/9}}.
\end{equation}

\subsubsection{DEJMPS Protocol}
\label{sec:deutsch-protocol}

The DEJMPS entanglement purification protocol, as initially proposed by its namesakes in Ref.~\cite{deutsch1996quantum}, is almost identical to the BBPSSW protocol, and differs only in the allowed input states and in the details of the local operations performed. Crucially, the central measure-and-exchange step (and thus the time spent communciating via the classical channel) is left unchanged. The two-qubit input states $\hat\rho^{[1(2)]}$ for the DEJMPS protocol need not be of Werner form, nor even drawn from the same ensemble; one could, for example, use DEJMPS to purify entanglement from two pairs of qubits received from two different SPDC sources which have taken entirely different routes through the network and have been subjected to entirely different types of noise. This fact makes it extremely useful in the context of noise models which do not restrict the quantum state to the Bell-diagonal (or indeed Werner) subspace. Similarly to the BBPSSW protocol, in order to obtain a probabilistic fidelity gain the \textit{average} fidelity must exceed $F \geq 1/2$.

Due to the above discussion, the DEJMPS protocol does not require twirling to bring the state to Werner form, and substitutes the initial local $Y$-rotation of BBPSSW with a more complicated operation $\hat R = u_1 \otimes u_1$ where $u_1$ is given by the deterministic twirling Kraus operator of \cref{eq:u1-kraus}. The complete pre-measurement local operation is given by
\begin{equation}
    \hat \rho_\mathrm{MX}(F) =  R_{AB} R_{A'B'}~\left(\hat\rho^{[1]}_{AB} \otimes \hat\rho^{[2]}_{A'B'}\right) ~ R^\dagger_{AB} R^\dagger_{A'B'},
    \label{eq:deutsch-rotation}
\end{equation}
where we have inserted explicit space subscripts, and where, again, the MX subscript denotes the fact that this is the final combined state of the entangled qubit pairs before the non-deterministic measurement step; the argument $F$ should be understood as the mean fidelity of the centers of the ensembles from which $\hat\rho_{1(2)}$ are drawn. The measure-and-compare steps proceed identically to BBPSSW, with success corresponding to matching measurement outcomes. On protocol success, the state is postprocessed by partial local $Y$-rotations according to \cref{eq:inverse-y-rot} identically to the BBPSSW protocol.

\begin{figure}[!ht]
    \centering
    \includegraphics[width=\linewidth]{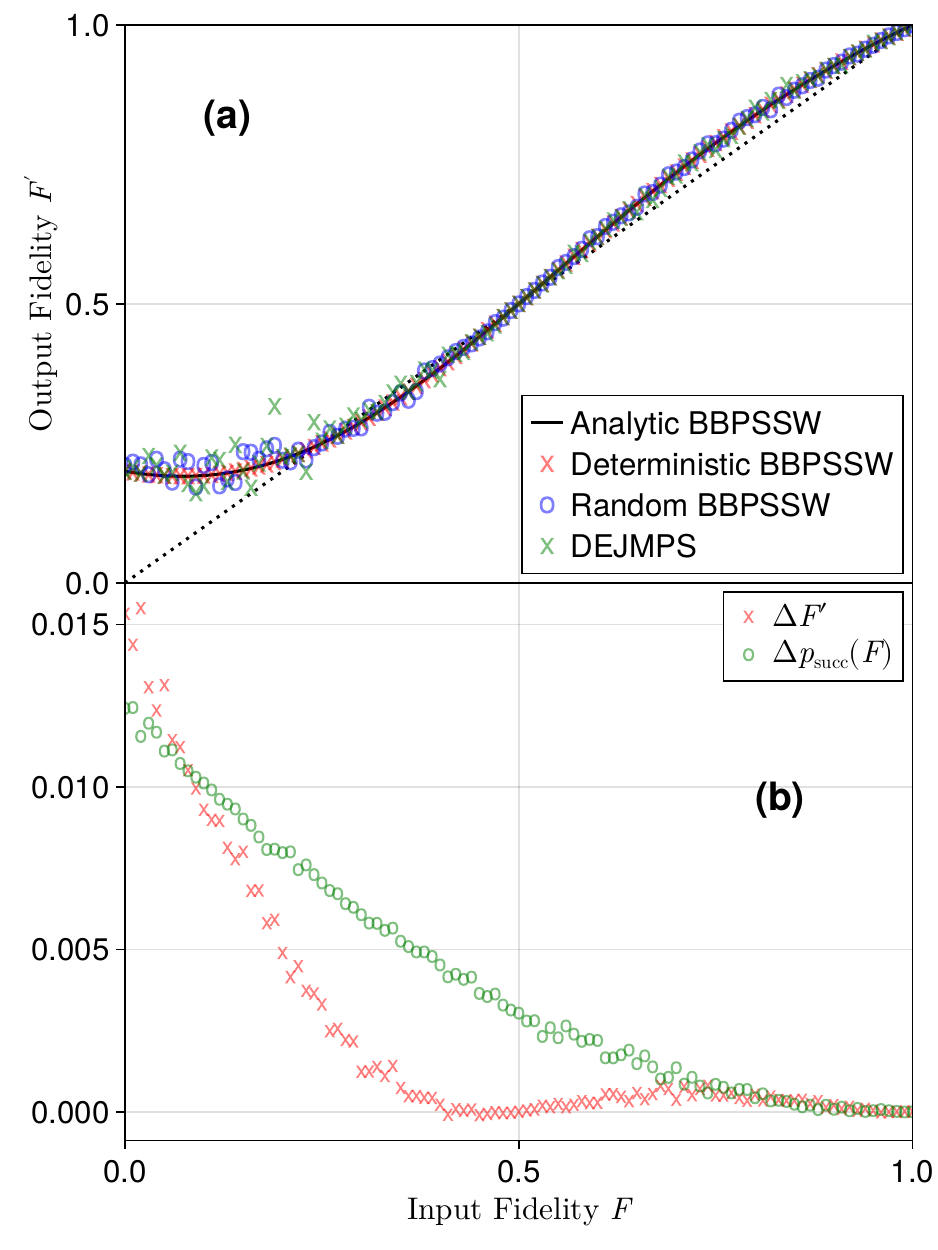}
    \caption{\textbf{Comparison of BBPSSW and DEJMPS protocols on random test states}. Panel \textbf{(a)} shows a direct comparison of fidelity gains $F'$ for initial states $\hat \rho_W(F)$ for BBPSSW protocols with deterministic and random twirlings and the DEJMPS protocol, against the analytic BBPSSW curve of \cref{eq:analytic-bennett} (solid black line) and the break-even point $F=F'$ (dotted black line), respectively. Panel \textbf{(b)} shows the differences in fidelity gain $\Delta F^\prime$ and success probability $\Delta p_\mathrm{succ}(F)$ between the DEJMPS and deterministic BBPSSW protocols, respectively (see \cref{eq:perfromance_diff} and discussion thereof). Non-deterministic curves in panel \textbf{(a)} averaged over a low number $2^3$ samples to illustrate fidelity gain spread, curves in panel \textbf{(b)} averaged over a high number $2^{15}$ samples to ensure convergence.}
    \label{fig:protocol-compare}
\end{figure}

\subsubsection{Comparison of BBPSSW and DEJMPS}
\label{sec:protocol-compare}

\begin{figure*}[ht!]
    \centering
    \includegraphics[width=\textwidth, height=\textheight, keepaspectratio]{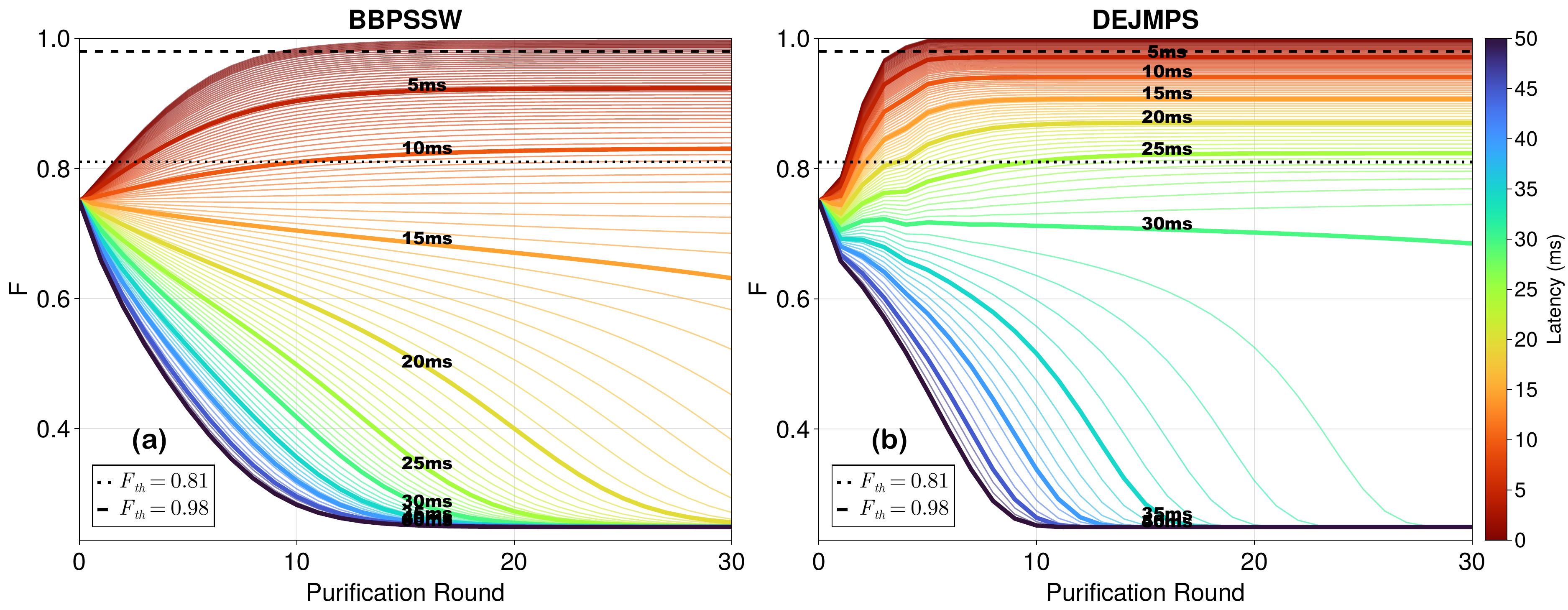}
    \caption{Color coded plots showing fidelity as a function of purification round and classical latency for both (a) BBPSSW and (b) DEJMPS. Results shown for the calcium ion–trap (\(^{40}\mathrm{Ca}^+\)) platform with quantum memory parameters \(T_1=1.14\,\mathrm{s}\), \(T_2=0.5\,\mathrm{s}\) (see \cref{tab:tab1}) and initial fidelity \(F_0=0.75\). Each colored curve corresponds to a distinct one-way classical latency \(T_C\); latencies are encoded by a continuous color scale spanning \(0\text{--}50\,\mathrm{ms}\) with values sampled from the empirical distribution in Fig.~\ref{fig:fidelity-fixed_latency}. Horizontal reference lines indicate the application thresholds for QKD \((F_{\mathrm{th}}=0.81)\) and DQC \((F_{\mathrm{th}}=0.98)\). Curves are plotted up to 30 rounds.}
    \label{fig:fidelity_rounds_ion_trap}
\end{figure*}

As previously noted in \cref{sec:deutsch-protocol} there are two main operational differences between the BBPSSW and DEJMPS protocols; firstly that the local DEJMPS operation $\hat R$ of \cref{eq:deutsch-rotation} is harder to implement than the simple $Y$-rotation of BBPSSW, and secondly that (because of this) the DEJMPS protocol does not impose any special constraints on the types of input state allowed (whereas input states to the BBPSSW protocol must be drawn from, or artificially brought into via twirling, identical ensembles of Werner form). Additionally, the DEJMPS protocol has a higher expected fidelity yield $F'$ than BBPSSW, with deterministic twirling leading to higher fidelity yield in BBPSSW than random twirling \cite{bennett1996mixed, deutsch1996quantum}. In \cref{fig:protocol-compare}\textbf{(a)} we present a numerical analysis of DEJMPS (green crosses), random BBPSSW (blue circles), and deterministic BBPSSW (red crosses) which all approximately follow the characteristic analytic BBPSSW formula of \cref{eq:analytic-bennett}, and which all exhibit fidelity gain $F' > F$ for $F > 1/2$ as expected. We illustrate the difference between the DEJMPS and the deterministic BBPSSW protocols by comparing them directly in \cref{fig:protocol-compare}. In panel \cref{fig:protocol-compare}\textbf{(a)} we plot the results of all three protocols (averaged over $2^3$ input states) and find that they all follow the analytic BBPSSW protocol curve (solid black line) of \cref{eq:analytic-bennett} closely, with fidelity gain $F' > F$ above the $F=F'$ curve (dotted black line) for ensemble fidelities above the $F > 1/2$ threshold. Notably, the  BBPSSW protocol with deterministic twirling (red crosses), exhibits much less spread than the random protocols (emphasized by our use of few samples in this panel) but achieves slightly lower output fidelities. In panel \cref{fig:protocol-compare}\textbf{(b)} we investigate the excess fidelity gain $\Delta F'$ and protocol success probability $\Delta p'_\mathrm{succ}(F)$ of the DEJMPS protocol relative to the random BBPSSW protocol; defined as:
\begin{align}
    \Delta F' &= F'_\mathrm{DEJMPS} - F'_\mathrm{BBPSSW}, \\ p'_\mathrm{succ}(F) &= p^\mathrm{DEJMPS}_\mathrm{succ}(F) - p^\mathrm{BBPSSW}_\mathrm{succ}(F).
    \label{eq:perfromance_diff}
\end{align}
We use a high number $2^{15}$ of samples to ensure convergence. As expected, we find that the DEJMPS protocol outperforms BBPSSW (albeit marginally) in both regards in the threshold region $F>1/2$. This marginal advantage snowballs dramatically in the context of actual operational resource usage, resulting in DEJMPS potentially outperforming BBPSSW by several orders of magnitude in terms of entangled pair throughput (see \cref{fig:all_rates} and discussion thereof in \cref{subsec:distillablerate}).

\subsection{Total Distillable Rates}
\label{subsec:ratelat}

The central question of this article is whether entanglement purification protocols can be enacted over realistic networks quickly enough to shield, or improve, the fidelity of qubits idling in memory. Essentially, information is lost during each round of entanglement purification during the exchange of classical information; in this step of the protocol the qubits are idling in memory and decohere according to the Lindblad equation such that the final purified state fidelity $F$ with the Bell singlet is reduced via Lindblad dynamics. Numerically this is realized in \cref{sec:results} by propagating the output state $\hat \rho(F')$ for the classical latency time under a Lindblad equation defined in \cref{sec:lindblad}, which is, in turn, defined by the $T_1$ and $T_2$ parameters for a given quantum memory. In this section we introduce the \textit{total distillable rate} $R(F_\mathrm{th})$: the rate at which entangled pairs satisfying $F \geq F_{\mathrm{th}}$ can be delivered by the purification protocol under Lindblad idle dynamics; and a key operational metric for quantum networks. If, for example, a sufficiently high distillable rate $R(F_\mathrm{th})$ can be achieved for the QKD threshold fidelity $F_\mathrm{th} = 0.81$, then the network can sustain QKD protocols across nodes $A$ and $B$ (similarly for, e.g., the DQC threshold $F_\mathrm{th} = 0.98$, or any other fidelity threshold).

Our formulation of $R(F_\mathrm{th})$ captures the long-time steady-state distillable rate for arbitrary threshold fidelities $F_\mathrm{th}$. While different algorithms have been proposed for the scheduling of the purification of entanglement (see e.g., Ref.~\cite{van2008system} and references therein), we here consider simple symmetric scheduling in which a purification attempt occurs only if the fidelities of the candidate input pairs match. In this setting, each round of purification (indexed by $i$) involves two states of equal fidelity $F_i$ and a corresponding purification success rate $p(F_i)$ (as quantified by \cref{eq:success-probability}); after $n$ such rounds one reaches or exceeds the threshold fidelity $F_n \geq F_\mathrm{th}$ and the protocol ends. The expected number of attempts required to successfully purify at a given round is $1/p(F_i)$, and each round requires two such successful attempts before the next round can be attempted. Thus, the total expected number of incoming pairs required to complete the entire protocol (i.e., all the rounds of purification necessary to combat decoherence and achieve a fidelity $F \geq F_\mathrm{th}$ above the threshold) is given by
\begin{equation}
    E(F_\mathrm{th}) = \prod_{i=1}^n \frac{2}{p(F_i)}.
    \label{eq:epc_expression}
\end{equation}

\begin{figure*}[ht!]
    \centering
    \includegraphics[width=\textwidth, height=\textheight, keepaspectratio]{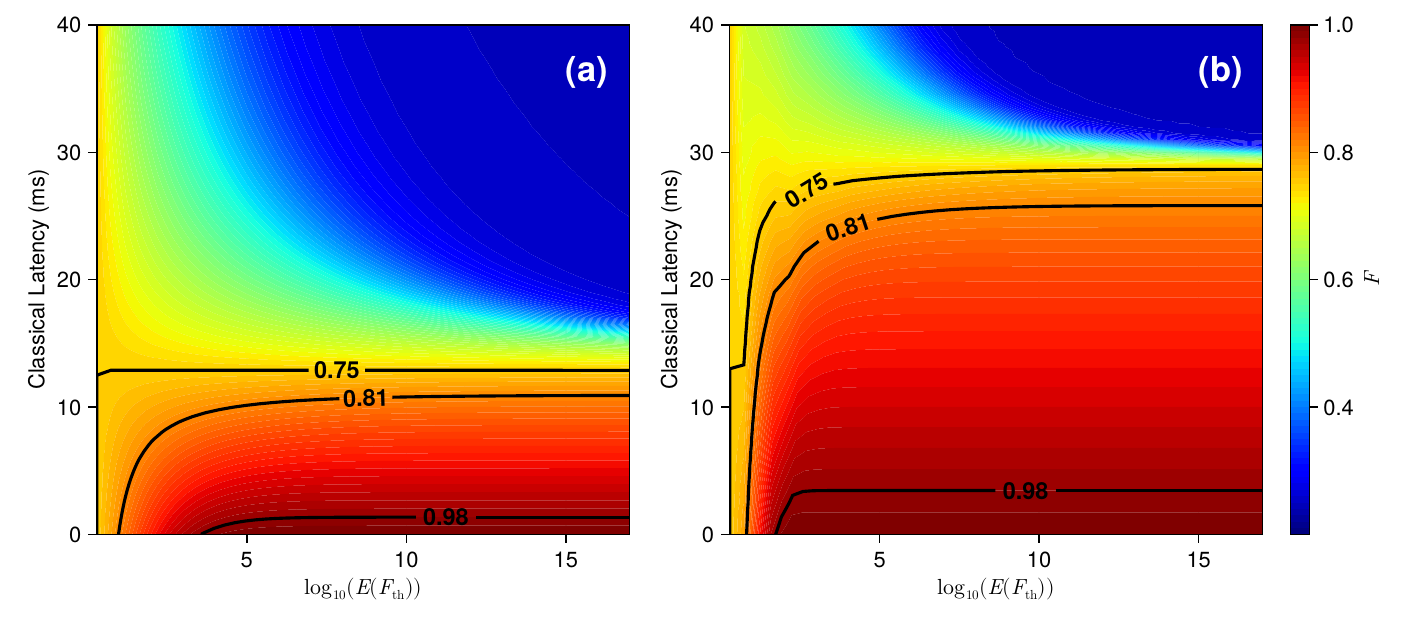}
    \caption{Heatmaps showing the final achievable Bell singlet fidelities $F$ as a function of the expected total number of consumed pairs $E$ and the classical latency for both \textbf{(a)} BBPSSW and \textbf{(b)} DEJMPS protocols respectively. Results shown for the Calcium ion trap (\(^{40}\text{Ca}^+\)) platform with quantum memory parameters $T_1 = 1.14$s, $T_2 = 0.5$s (see \cref{tab:tab1} and discussion thereof) and initial fidelity $F_0 = 0.75$. Black iso-fidelity curves show the break-even curve $F=F_0$, and the boundaries delineating feasible QKD ($F=0.81$) and DQC ($F=0.98$) respectively. All data points are averaged over $1024$ random initial states.}
    \label{fig:epc_consumption}
\end{figure*}

Assuming that base-pair generation takes place at a steady rate $R_{\mathrm{pair}}$ between the quantum nodes (determined by, inter alia, source production rate, loss in the fibers, and flying to stationary qubit conversion success rates), and that the system can immediately initiate purification once the required pairs are available, the long-term (steady-state) throughput is:
\begin{equation}
  R(F_\mathrm{th}) = \frac{R_\mathrm{pair}}{E(F_\mathrm{th})}.
  \label{eq:distillable_rate}
\end{equation}
We finally remark that while $E(\cdot)$ is determined entirely by $F_0$, $F_\mathrm{th}$, the quantum memories,  classical network latencies, and choice of purification protocol, it is more useful as a user-controlled parameter. In this setting, $R(F_\mathrm{th})$ can be tuned by varying $E(F_\mathrm{th})$ manually, and determining the attainable values of $F_\mathrm{th}$ in the context of a given architecture. We carry out such an analysis in \cref{sec:results} for attainable threshold fidelities $F_\mathrm{th}$ as a function of classical latencies and the required resource overhead $E$, and map the distinct fidelity-gain and fidelity-loss regimes, highlighting the specific iso-fidelity curves that define the operational requirements for QKD ($F_\mathrm{th} \geq 0.81$) and DQC ($F_\mathrm{th} \geq 0.98$) thresholds, respectively.  We then leverage this analysis into an investigation of whether or not these fidelities (and efficient entanglement purification in general) are attainable in current-generation and near-future network architectures.


\section{Results}
\label{sec:results}

Here we numerically investigate the performance of the BBPSSW and DEJMPS entanglement purification protocols across several of network latencies and quantum memory coherence times $(T_1,\,T_2)$ listed in \cref{tab:tab1}. We use the ion-trap (\(^{40}\mathrm{Ca}^+\)) platform with $T_1 = 1.14\mathrm{s}$ and $T_2 = 0.5\mathrm{s}$ as our primary benchmark. All simulations are performed in Julia, with the initial Bell singlet fidelity set to $F_0=0.75$ to reflect a highly conservative worst‐case scenario that accounts for the significant storage decoherence caused by tail latencies in the initial pair verification step, channel noise during fiber transmission, and errors from the quantum transduction process in memory. 

The results are organized into three interconnected parts that link hardware constraints to the level of service a quantum network can realistically deliver. In \cref{subsec:rounds}, we analyze how entanglement fidelity evolves through successive rounds of purification when qubits undergo decoherence during classical communication delays. This establishes the feasible operating regime of the network: for given memory coherence times ($T_1$, $T_2$), latency characteristics of the classical channel, and initial fidelity $F_0$, we can determine whether purification is feasible, whether it produces entanglement that meets application-specific fidelity thresholds (e.g., 0.81 for QKD or 0.98 for distributed quantum computing), and the total throughput of entangled pairs which meet these fidelity thresholds. Then, in \cref{subsec:epc}, we assess the resource demands for meeting these fidelity thresholds via the expected entangled-pair consumption $E(F_\mathrm{th})$ of \cref{eq:epc_expression}, reframing the more abstract notion of feasibility as the practical operational question of how many noisy input pairs must be consumed to reach the threshold fidelity. Finally, in \cref{subsec:distillablerate} we evaluate the achievable distillable entanglement rate $R(F_\mathrm{th})$ of \cref{eq:distillable_rate}, representing the steady-state throughput of high-fidelity pairs deliverable under a given network configuration. By comparing certain well-known fidelity bounds with the resource constraints mapped out in \cref{subsec:rounds} and \cref{subsec:epc}, we provide a direct assessment of the network’s service capability: the rate at which sufficiently high-fidelity entangled pairs can be sustained to support quantum applications under realistic operating conditions.

\subsection{Fidelity Improvement per Round}
\label{subsec:rounds}

The central object of our inquiry is the trade-off between the fidelity gained through with successive purification rounds, and the fidelity lost due to memory decoherence induced by classical communication delays during the implementation of said purification protocols. \cref{fig:fidelity_rounds_ion_trap} illustrates this behavior for the BBPSSW and DEJMPS protocols, respectively for ion-trap (\(^{40}\text{Ca}^+\)) platform ( corresponding results for the rare-earth ion and NV-center platforms are provided in \cref{sec:fidelityvspurification}) that displays families of fidelity trajectories across the number of purification rounds. Each curve corresponds to distinct classical wait times per round, taken as the maximum of the two directional one-way latency, and latencies are encoded by a continuous color scale spanning $0-50$ms, with values sampled from the empirical distribution in \cref{fig:fidelity-fixed_latency}. Horizontal reference lines mark the application thresholds $F_{th} = 0.81$ (QKD) and $F_{th} = 0.98$ (DQC). Trajectories start from $F_{0} = 0.75$ and are shown up to 30 rounds.

\begin{figure*}[!t]
  \centering
  \includegraphics[width=\linewidth, height=\textheight, keepaspectratio]{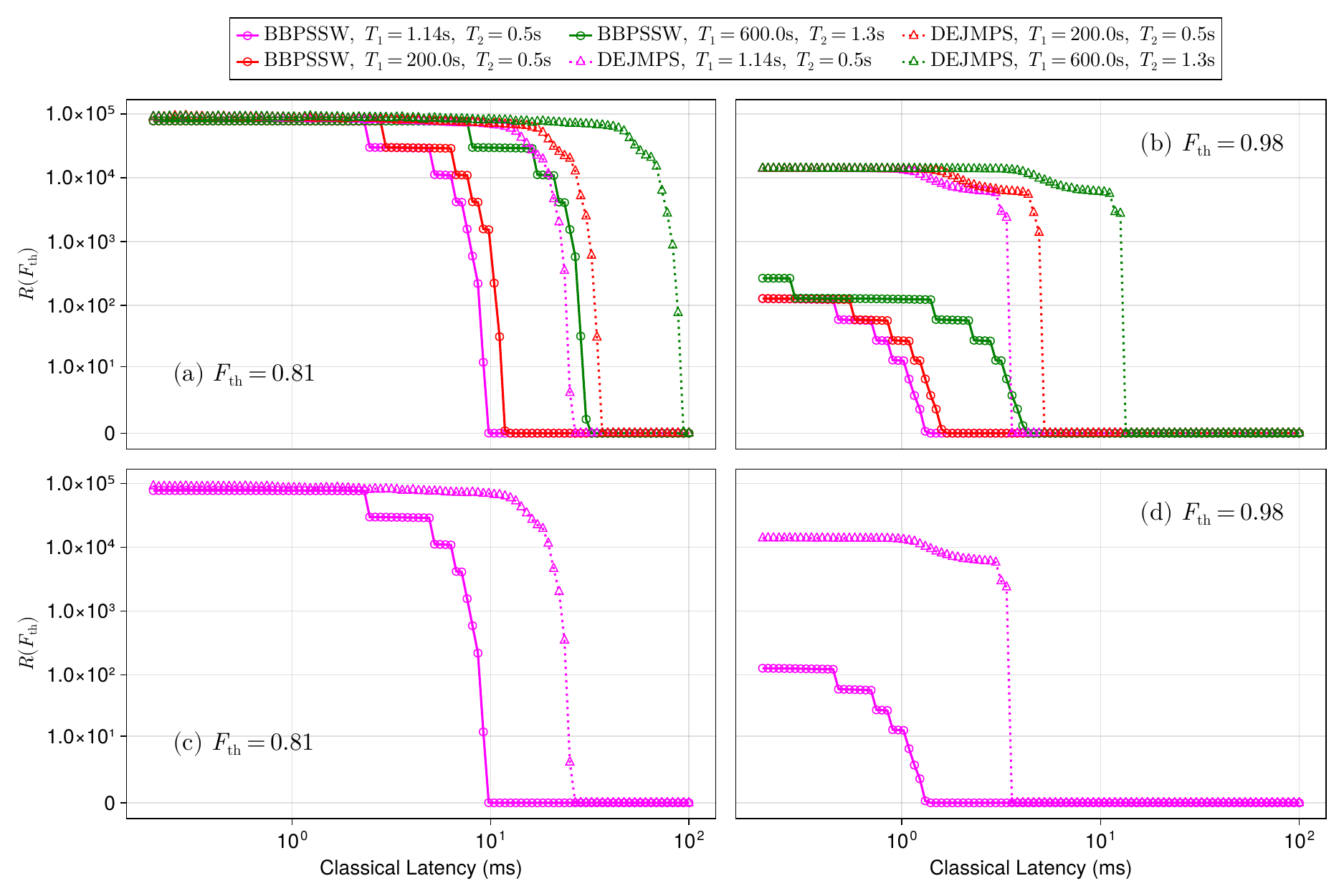}
  \caption{Distillable rate $R(F_\mathrm{th})$ as a function of the classical latency. All panels show results for both BBPSSW (solid line with circle markers) and DEJMPS (dotted line with triange markers). Panel \textbf{(a)} shows results for the threshold fidelity $F_\mathrm{th}=0.81$ required for QKD, and for a variety of different $T_1$ and $T_2$ times informed by the quantum memories listed in \cref{tab:tab1}. Panel \textbf{(b)} shows similar results for the threshold fidelity $F_\mathrm{th}=0.81$ required for DQC. Panels \textbf{(c)} and \textbf{(d)} isolate the results for the Calcium ion trap (\(^{40}\text{Ca}^+\)) platform from panels \textbf{(a)} and \textbf{(b)} such that they can be more clearly compared. The Calcium ion trap memory parameters $T_1 = 1.14$s, $T_2 = 0.5$s, are used as the baseline parameters throughout this article. All panels show an initial fidelity $F_0 = 0.75$, and all data points are averaged over $1024$ random initial states.}
  \label{fig:all_rates}
\end{figure*}

Qualitatively, small latencies yield steep rises that converge to high steady-state fidelities;  as latency increases, the initial gain is damped and the plateau drops. In the high-latency regime, trajectories collapse toward a noise-dominated fixed point, approaching the maximally mixed state with fidelity \(F \simeq \tfrac{1}{4}\). Throughout, DEJMPS lies above BBPSSW and reaches its plateau in fewer rounds.
A key feature of the plot is the identification of latency ranges where either (i) decoherence offsets purification gains to such a degree that the protocol fails entirely, and the steady-state fidelity approaches the maximally mixed value of $F=1/4$; and (ii) decoherence offsets limit the steady-state value of the fidelity such that it cannot exceed the relevant application threshold $F_\mathrm{th}$. For distributed quantum computing (DQC) the threshold fidelity is $F_{th} = 0.98$; which both BBPSSW and DEJMPS satisfy only within the lowest latency band $0-5$ms. The crossover to $F < 0.98$
occurs within this band for each protocol, and for latencies in the $5-10$ms band and above, neither can sustain DQC-grade fidelity, even after multiple purification rounds. However, DEJMPS achieves marginally higher fidelities within the $0-5$ms band and requires fewer rounds to converge to the threshold fidelity.

For the more relaxed quantum key distribution (QKD) threshold $F_{th} = 0.81$, the latency range over which the maximum fidelity (over rounds) remains at or above the threshold is broader. BBPSSW maintains fidelities above 0.81 through the $5-10$ms band but drops below threshold in the $10 - 15$ms interval. DEJMPS performs more robustly, remaining above 0.81 up to the $20-25$ms band before crossing below within the $25-30$ms range. At latencies beyond approximately $30$ms, both protocols plateau below the QKD requirement, and additional purification rounds do not yield any further improvement.

\subsection{Entangled pairs consumption for producing high entangled fidelity pair for various latencies}
\label{subsec:epc}

The per-round fidelity analysis (\cref{fig:fidelity_rounds_ion_trap}) identified, as a function of classical latency, the regimes in which purification either (i) fails generally, or (ii) fails to increase the fidelity via successive purification rounds to above certain application thresholds. In this section, we perform a more systematic interrogation of these features in the context of the cost to the network in quantum resources. In doing so we turn to the total number of consumed pairs $E(F_\mathrm{th})$ introduced in \cref{eq:epc_expression}, which quantifies the expected total number of entangled pairs the network must consume in order to produce an entangled pair exceeding the threshold fidelity $F_\mathrm{th}$. We show a heatmap of the output fidelity $F$ against $E$ and the classical latency for the ion-trap (\(^{40}\mathrm{Ca}^+\)) platform (using the same memory parameters as used in \cref{subsec:rounds}) in \cref{fig:epc_consumption} (BBPSSW in panel \textbf{(a)} and DEJMPS in panel \textbf{(b)}). For a fixed latency, moving right along the $E$ axis bears the operational interpretation of allocating more base pairs to the protocol; either to successfully complete more rounds of repeated purification, or to complete rounds with lower success probabilities $p_\mathrm{succ}(F)$. 

The most striking feature of both panels of \cref{fig:epc_consumption} is the large blue region at $F=1/4$. This region represents the total failure of purification, and the degradation of the entangled pair to the (useless) maximally mixed state. Any trajectory through the plane towards this region corresponds to a situation in which decoherence offsets purification gains entirely, and successive purification rounds end with lower fidelity than the previous round. In network configurations with network latencies and source entangled pair rates within these region, it is simply better to use entangled pairs as they become available rather than attempting purification. We identify this region via the break-even iso-fidelity contour $F=F_0=0.75$, which divides the network configurations than can sustain or exceed the initial fidelity $F_0$ from those which cannot. Consistent with the earlier findings, the break-even contour for BBPSSW demarcates a larger no-gain region ($F < F_{0}$) than DEJMPS, indicating a broader regime in which purification yields no net benefit under practical resource budgets.

A second feature of \cref{fig:epc_consumption} is that fidelity changes rapidly at first and then saturates, revealing a diminishing-returns frontier beyond which additional consumption yields little or no gain as memory-induced decoherence cancels out the gains from any given purification round. On the heatmaps these appear as iso-fidelity contours that flatten with increasing $E$; in the per-round curves the same effect manifests as late-round flattening in \cref{fig:fidelity_rounds_ion_trap}, for $F_{th} = 0.98$, within the $0-5$ms band for both DEJMPS and BBPSSW, and, for $F_{th} = 0.81$, within the $5-10$ms band for BBPSSW and the $25-30$ms band for DEJMPS. Conversely, for a fixed fidelity threshold (following an iso-contour), moving upward in latency shifts the contour to the right: maintaining the same fidelity $F$ at higher latency requires substantially more $E$. In particular, the DQC contour (\(F{=}0.98\)) remains confined to the low-latency/low-$E$ corner, whereas the QKD contour (\(F{=}0.81\)) extends to much higher latencies, albeit at progressively larger $E$.

Practically, these plots provide a simple recipe for specifying system requirements, or determining whether an existing network configuration can sustain enough entanglement to succeed at a given application. In the first case one must simply define a threshold fidelity $F_\mathrm{th}$, and the system requirements in terms of network latencies and base pair production rates can be read directly off \cref{fig:epc_consumption}. In the second case, one is given a latency band and pair production count $R$, and identifies the entire region to the left of $E=R$ as the operating regime of the network. If no iso-fidelity curve at a threshold value $F_\mathrm{th}$ intersects with this region then said threshold is not attainable at that latency/protocol on the given hardware. Whilst we have carried this numerical analysis out for a specific set of input parameters $T_1$, $T_2$, and $F_0$; the calculations are not too numerically intensive and can in principle be carried out for any pair of nodes in a given network.

\subsection{Distillable entanglement rate}
\label{subsec:distillablerate}

Finally, we use a service level metric $R(F_{\mathrm{th}})$, the distillable rate defined in \cref{eq:distillable_rate} to capture end-to-end network performance under given latency and hardware constraints that reflects the throughput of high-fidelity states available to applications. The network topology used for the simulation is depicted in \cref{fig:schematic}. For the quantum network, as per the node architecture presented in \cite{bali2025routing}, the loss due to traversing an intermediate node is considered to be 8 dB (e.g., due to optical wavelength switching equipment) and the loss due to the source node and the memory node is considered to be 4 dB (e.g., wavelength add/drop losses). Fiber attenuation is considered to be 0.2 dB/km. The entangled‐photon generation rate at the source is set to 1.3\,MHz~\cite{sansa2022visible}.  We carry out this study for the $A$-$B$ node pair. For a given base pair production rate $R_\mathrm{pair}$ at nodes $A$ and $B$ (after accounting for losses mentioned above), $R(F_{\mathrm{th}}) = R_\mathrm{pair}/E(F_\mathrm{th})$ gives the expected steady-state rate at which entangled pairs above the threshold fidelity can be produced between nodes $A$ and $B$; see \cref{eq:distillable_rate} for details.

In \cref{fig:all_rates} we show the distillable rate $R(F_{\mathrm{th}})$ plotted against classical latency (log scale) across memory platforms for both BBPSSW and DEJMPS. Panels \textbf{(a)} and \textbf{(c)} correspond to the QKD threshold (\(F_{\mathrm{th}}{=}0.81\)), while \textbf{(b)} and \textbf{(d)} correspond to the DQC threshold (\(F_{\mathrm{th}}{=}0.98\)). The most striking feature of all figures is that, whilst the distillable rate falls off generally with increasing latency, it suddenly collapses towards zero at some critical latency, beyond which the threshold fidelity is no longer attainable. This is in line with the results discussed in \cref{subsec:rounds} and \cref{subsec:epc}. Three additional features can also be identified. 

First, we note a clearly visible gap between the results for BBPSSW and DEJMPS protocols across all panels \textbf{(a)-(d)} for each fixed platform and threshold. In all cases the DEJMPS curves lie above the corresponding BBPSSW curves at the same latency often exceeding the BBPSSW distillable rate by several orders of magnitude in the lower latency decades. This is an expected feature of DEJMPS which generally yields higher fidelity gains and success probabilities than BBPSSW (see \cref{fig:protocol-compare}), and which thus converges to threshold fidelities faster and with a lower pair consumption rate. This gap narrows as the curves approach the zero-rate region.

Second, for fixed protocol and thresholds, panels \textbf{(a)} (QKD) and \textbf{(b)} (DQC) indicate that longer-coherence platforms shift the positive-rate region to higher latency decades: rare-earth ions sustain nonzero rates over the widest span, NV centers are intermediate, and ion traps are the most constrained. This ordering is consistent with the crossover latency bands observed in \cref{fig:fidelity_rounds_ion_trap} (and with the supplementary results for rare-earth \cref{fig:fidelity_rounds_rare_earth} and NV-center \cref{fig:fidelity_rounds_nv_center} platforms shown in \cref{sec:fidelityvspurification}), and with common sense: using quantum memories with longer decoherence times should improve the impact of successive rounds of purification as less decoherence occurs during classical communication.

Finally, comparing panel \textbf{(c)} with panel \textbf{(d)} clearly demonstrates the systematic gap between the higher distillable rates attainable for lower-fidelity thresholds against the lower distillable rates attainable for higher-fidelity thresholds. In the first latency decade \(( 10^{0}\!-\!10^{1}\mathrm{ms})\) 
, \(R(0.81)\) exceeds \(R(0.98)\) by about one order of magnitude for DEJMPS and by roughly four orders for BBPSSW. For \(F_{\mathrm{th}}=0.81\), DEJMPS achieves \(R(0.81)\approx \!10^{5}\) pairs/s in the first decade \((10^{0}\!-\!10^{1}\mathrm{ms})\) and maintains roughly the same rate throughout this decade then tapers to zero early in the next \((10^{1}\!-\!10^{2}\mathrm{ms})\) decade; BBPSSW begins the decade \((10^{0}\!-\!10^{1}\mathrm{ms})\) with roughly similar rate of \(R(0.81)\approx \!10^{5}\) pairs/s, which then declines, approaching zero near the \(\!10^{1}\mathrm{ms}\) mark. For \(F_{\mathrm{th}}=0.98\), the positive rates of both protocols are confined to the first decade \((10^{0}\!-\!10^{1}\mathrm{ms})\). The BBPSSW protocol starts the decade with $R(0.98)\approx 10^{1}$ pairs/s and collapses almost immediately. In contrast, the DEJMPS protocol begins the decade with a much higher rate of $R(0.98)\approx  10^{4}$ pairs/s and maintains a positive rate for slightly longer before falling to zero by the middle of this decade. This feature is due to the tapering off of the fidelity gain $F'$ as $F \to 1$ for both BBPSSW and DEJMPS, which ensures that high-fidelity thresholds are hard to attain, and are exponentially expensive in terms of consumed pairs. Operationally, this implies that even if a high fidelity threshold is theoretically attainable, it may be practically infeasible due the exponentially large number of entangled pairs required to produce it. 

Practically, the non-zero portions of all curves are delimited by the same iso-fidelity contours highlighted in \cref{fig:epc_consumption}: as the classical latency crosses a given contour, purification can no longer attain the corresponding threshold fidelity \(F_{\mathrm{th}}\) and \(R(F_{\mathrm{th}})\!\to\!0\). Approaching break-even from below, \(R(F_{\mathrm{th}})\) remains non-zero but decreases with increasing latency; at the band edge it collapses to zero as repeated purification suddenly fails. Beyond this value the threshold fidelity is unattainable and the distillable rate vanishes. This shared structure across panels (a–d) yields a clear design rule: select protocol and hardware so that the operating point remains within the iso-fidelity contour at the required threshold \(F_{\mathrm{th}}\).

\section{Conclusions}
\label{sec:conclusion}

This work investigates the challenge of implementing entanglement purification in quantum networks with the practical constraint in which protocol gains are in direct competition with decoherence induced by classical communication latency. Our analysis moves beyond idealized assumptions to provide a quantitative framework for assessing the practical viability of purification. We have shown that the interplay between protocol choice, quantum memory quality, and classical network latency defines sharp operational boundaries that determine whether purification provides a net benefit or is actively detrimental.

The primary contribution of this research is an engineering guideline that translates these fundamental trade-offs into actionable design rules. By mapping the achievable fidelity against resource consumption and latency, we allow network architects to determine the necessary system parameters needed to meet application-specific fidelity thresholds for QKD or DQC, including latency budgets for the classical control plane, minimum coherence time requirements for quantum memories, and the expected resource overhead in terms of initial Bell pairs.

Our results also quantify the consistent and significant performance advantage of the DEJMPS protocol over BBPSSW under realistic conditions, establishing it as a more robust choice for deployment. Ultimately, this work provides a foundational methodology for engineering large-scale quantum networks, bridging the gap between abstract protocols and the physical-layer realities of their implementation.

An immediate future extension of this work is the exploration of different protocols, or more advanced entanglement purification protocols such as hashing and breeding. These schemes are asymptotically more efficient than BBPSSW or DEJMPS, operating on large ensembles of entangled pairs jointly rather than pairwise. In principle, hashing and breeding can achieve a nonzero yield of distilled high-fidelity pairs in the limit of many inputs, whereas iterative two-pair methods consume a large fraction of pairs and see their yield vanish as threshold fidelity increases. However, there are significant challenges facing such protocols in practice, notably their extreme sensitivity to local noise; with hybrid approaches presenting a promising middle ground \cite{zwerger2014robustness}. Another possible extension involves the analysis of how alternative entanglement purification scheduling, beyond the symmetric scheduling considered here, influence performance in terms of fidelity, resource consumption and throughput \cite{van2008system}. Finally, one could consider non-Markovian local memory effects or inter-pair entanglement generation in the memories, to further refine the microscopic treatment of the dynamical evolution of entangled states in the network. In any of the above settings, our work introduces mathematical methods and operational perspectives which place the future analysis of trade-offs between protocols and the environment on a firm foundation.

\section*{Acknowledgment}
This material is based upon work supported by the Science Foundation Ireland grants 20/US/3708, 21/US-C2C/3750, and 13/RC/2077 P2 and National Science Foundation under Grant No. CNS-2107265.



\appendix 

\subsection{Lindblad Integrator Convergence Checks}
\label{sec:convergence}

\begin{figure}[ht!]
    \centering
    \includegraphics[width=\linewidth]{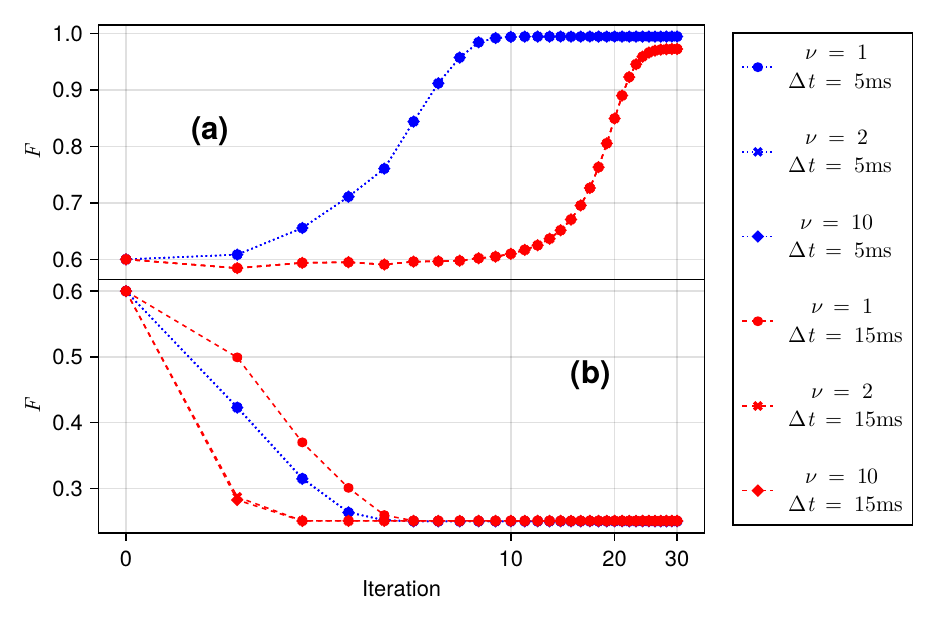}
    \caption{Numerical evidence for convergence of the $\nu$-step fourth-order Runge-Kutta integrator for the Lindblad equation for \textbf{(a)} the parameters used in the main text $T_1 = 1.14$s, $T_2 = 0.5$s, and \textbf{(b)} $T_1 = 1.14$s$/20$, $T_2 = 0.5$s$/20$.}
    \label{fig:convergence-checks}
\end{figure}

We integrate the Lindblad equation of \cref{eq:lindblad} using a fourth order Runge-Kutta scheme where the latency time $\Delta t$ is subdivided into $\nu$ steps of size $\delta t = \Delta t/\nu$. Given the trivial unitary dynamics, and the relative smallness of the latency times compared to the $T_1$ and $T_2$ times we investigate in the main text, it is reasonable to assume that the single-step $\nu = 1$ integrator is sufficiently accurate. We validate this in \cref{fig:convergence-checks}\textbf{(a)} for the parameters $T_1 = 1.14$s, $T_2 = 0.50$s used in the main text; in which we find that the $\nu \in \{1,2,10\}$ integrators all converge across many iterations of Lindblad decoherence interspersed with the DEJMPS purification protocol for latencies of both $\Delta t = 5$ms (blue curves) and $\Delta t = 15$ms (red curves). We contrast these checks to panel \textbf{(b)} wherein both parameters have been dramatically reduced by a factor of $20$, $T_1 = 1.14$s$/20$, $T_2 = 0.5$s$/20$; and where we still see convergence for sufficiently small latency times $\Delta t = 5$ms, but clear divergence of the $\nu = 1$ single-step integrator at the larger latency $\Delta t = 15$ms. These checks support the basic intuition that the $\nu=1$ single-step fourth order Runge-Kutta integrator suffices to accurately capture the physics interrogated in the main text.

\subsection{Fidelity Improvement per Round: Additional Platforms}
\label{sec:fidelityvspurification}

While the discussion in \cref{subsec:rounds} focuses on the ion-trap platform as a representative case, the same analysis was performed for rare-earth ions and NV centers. These additional results provided here, allow for cross-platform comparison of break-even latencies and highlight the role of memory coherence times in purification performance.

\subsubsection{Rare-Earth Ions (\(^{167}\)Er\(^{3+}\))}

\begin{figure*}[ht!]
    \centering
    \includegraphics[width=\textwidth, height=\textheight, keepaspectratio]{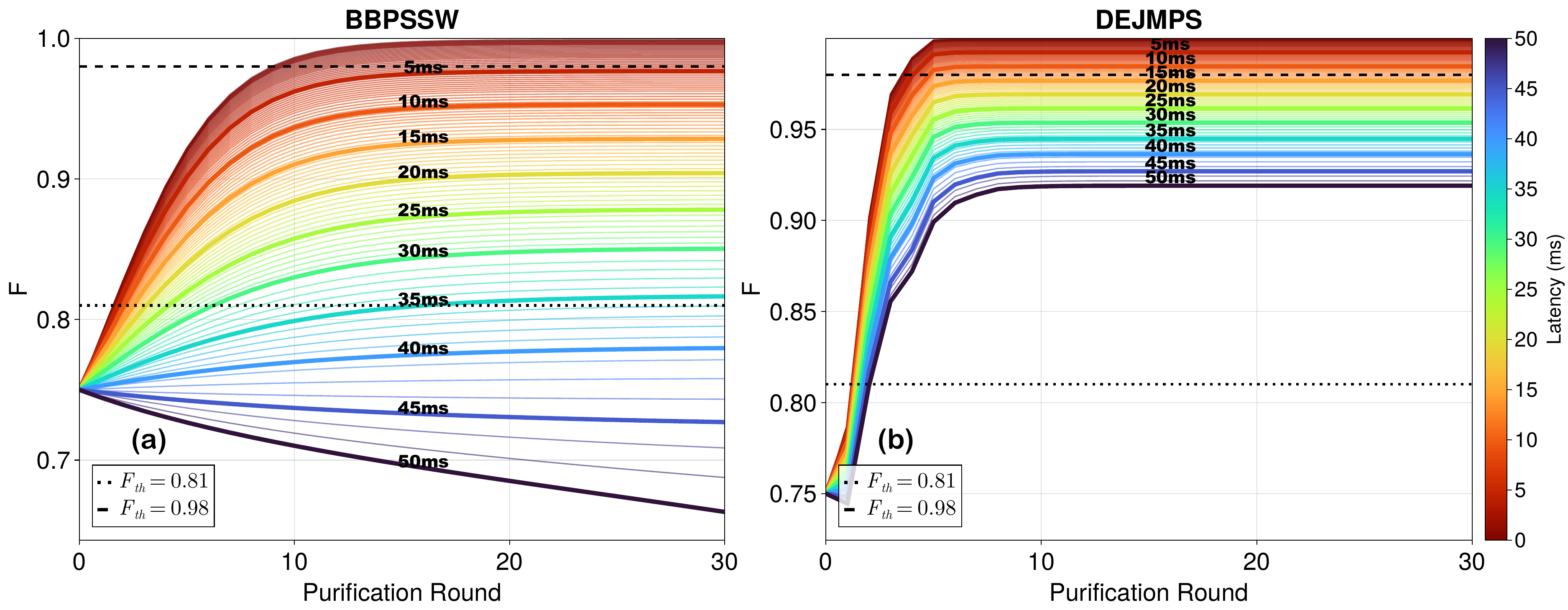}
    \caption{Color coded plots showing fidelity as a function of purification round and classical latency for both (a) BBPSSW and (b) DEJMPS. Results shown for the rare-earth ion (\(^{167}\mathrm{Er}^{3+}\)) platform with quantum memory parameters \(T_1=600\,\mathrm{s}\), \(T_2=1.3\,\mathrm{s}\) (see \cref{tab:tab1}) and initial fidelity \(F_0=0.75\). Each colored curve corresponds to a distinct one-way classical latency \(T_C\); latencies are encoded by a continuous color scale spanning \(0\text{--}50\,\mathrm{ms}\) with values sampled from the empirical distribution in Fig.~\ref{fig:fidelity-fixed_latency}. Horizontal reference lines indicate the application thresholds for QKD \((F_{\mathrm{th}}=0.81)\) and DQC \((F_{\mathrm{th}}=0.98)\). Curves are plotted up to 30 rounds.}
    \label{fig:fidelity_rounds_rare_earth}
\end{figure*}

For the rare-earth ion memory platform (
$T_1 = 600,  T_2 = 1.3 s$) the long coherence times significantly expand the range of tolerable classical latencies as evident in \cref{fig:fidelity_rounds_rare_earth}. The DQC threshold ($F_{th} = 0.98$) is sustained only within the $0-5$ms latency band for BBPSSW, whereas DEJMPS maintains $F > 0.98 $ until the $5-10$ms band before dropping below $F_{th}$ somewhere within $10-15$ms band. For the QKD threshold ($F_{th} = 0.81$), BBPSSW remains viable until the $30-35$ms band before crossing below within the $35-40$ms range, while DEJMPS maintains $F > 0.81$ across the full $0-50$ms latency spectrum. These results highlight the advantage of rare-earth memories in extending QKD feasibility to much larger latencies compared to the other platforms considered.

\subsubsection{NV Centers in Diamond}

\begin{figure*}[ht!]
    \centering
    \includegraphics[width=\textwidth, height=\textheight, keepaspectratio]{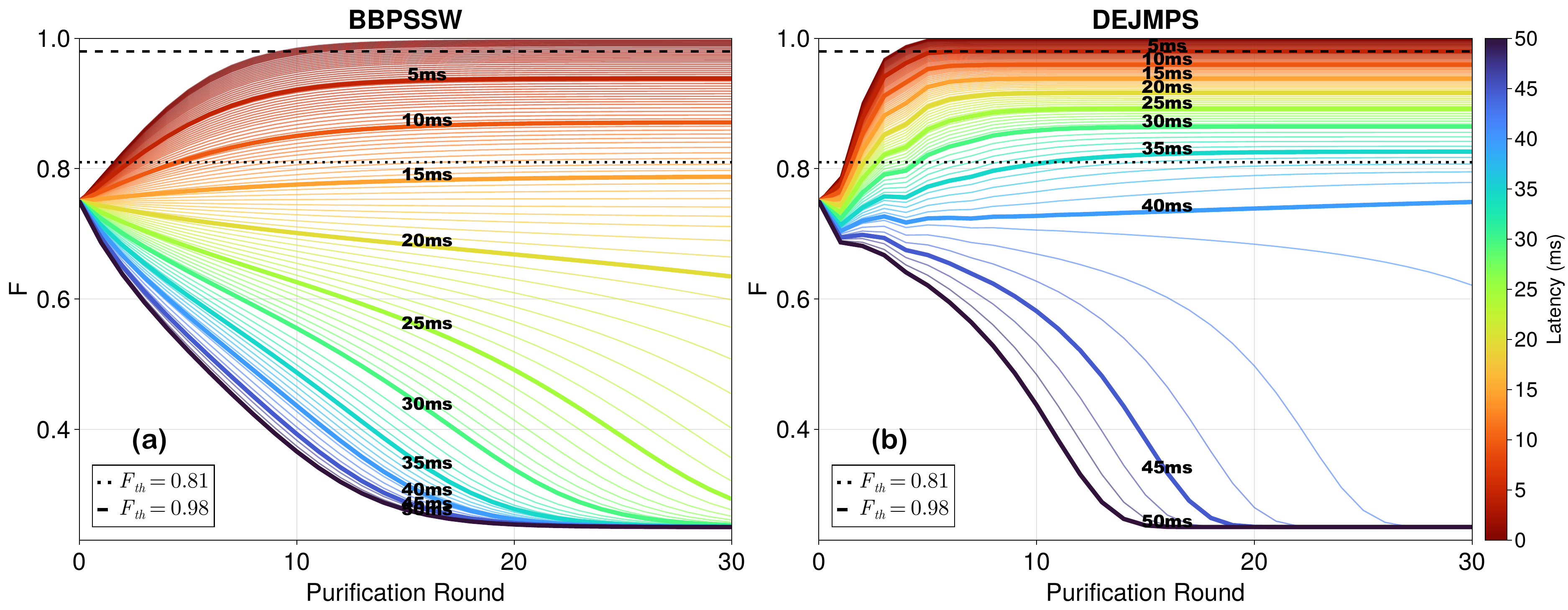}
    \caption{Color coded plots showing fidelity as a function of purification round and classical latency for both (a) BBPSSW and (b) DEJMPS. Results shown for NV centers in diamond with quantum memory parameters \(T_1=200\,\mathrm{s}\), \(T_2=0.5\,\mathrm{s}\) (see \cref{tab:tab1}) and initial fidelity \(F_0=0.75\). Each colored curve corresponds to a distinct one-way classical latency \(T_C\); latencies are encoded by a continuous color scale spanning \(0\text{--}50\,\mathrm{ms}\) with values sampled from the empirical distribution in Fig.~\ref{fig:fidelity-fixed_latency}. Horizontal reference lines indicate the application thresholds for QKD \((F_{\mathrm{th}}=0.81)\) and DQC \((F_{\mathrm{th}}=0.98)\). Curves are plotted up to 30 rounds.}
    \label{fig:fidelity_rounds_nv_center}
\end{figure*}

For the NV-center platform (
$T_1 = 200,  T_2 = 0.5 s $ ) we observe distinct break-even points for each application threshold. For the DQC threshold ($F_{th} = 0.98$), BBPSSW remains viable only in the $0-5$ms band, with the crossover occurring inside this interval. DEJMPS extends DQC feasibility slightly further, staying above 0.98 until the $5-10$ms band before dropping below threshold earlier within this band. For the QKD threshold ($F_{th} = 0.81$), BBPSSW remains above threshold until the $5-10$ms band and crosses below within $10-15$ms range, whereas DEJMPS maintains
$F > 0.81$ through the $30-35$ms band, with the crossover point appearing earlier inside the $35-40$ms interval. At latencies beyond ~$40$ms, both protocols level off below QKD-grade fidelity.

\bibliographystyle{IEEEtran}
\bibliography{references}

\begin{thebibliography}{10}
\providecommand{\url}[1]{#1}
\csname url@samestyle\endcsname
\providecommand{\newblock}{\relax}
\providecommand{\bibinfo}[2]{#2}
\providecommand{\BIBentrySTDinterwordspacing}{\spaceskip=0pt\relax}
\providecommand{\BIBentryALTinterwordstretchfactor}{4}
\providecommand{\BIBentryALTinterwordspacing}{\spaceskip=\fontdimen2\font plus
\BIBentryALTinterwordstretchfactor\fontdimen3\font minus \fontdimen4\font\relax}
\providecommand{\BIBforeignlanguage}[2]{{%
\expandafter\ifx\csname l@#1\endcsname\relax
\typeout{** WARNING: IEEEtran.bst: No hyphenation pattern has been}%
\typeout{** loaded for the language `#1'. Using the pattern for}%
\typeout{** the default language instead.}%
\else
\language=\csname l@#1\endcsname
\fi
#2}}
\providecommand{\BIBdecl}{\relax}
\BIBdecl

\bibitem{van2014quantum}
R.~Van~Meter, \emph{Quantum Networking}.\hskip 1em plus 0.5em minus 0.4em\relax Hoboken, NJ, USA: Wiley, 2014.

\bibitem{dur2007entanglement}
W.~D{\"u}r and H.~Briegel, ``Entanglement purification and quantum error correction,'' \emph{Rep. Prog. Phys.}, vol.~70, no.~8, p. 1381, 2007.

\bibitem{bennett1996purification}
C.~H. Bennett, G.~Brassard, S.~Popescu, B.~Schumacher, J.~A. Smolin, and W.~K. Wootters, ``Purification of noisy entanglement and faithful teleportation via noisy channels,'' \emph{Phys. Rev. Lett.}, vol.~76, no.~5, p. 722, 1996.

\bibitem{deutsch1996quantum}
D.~Deutsch, A.~Ekert, R.~Jozsa, C.~Macchiavello, S.~Popescu, and A.~Sanpera, ``Quantum privacy amplification and the security of quantum cryptography over noisy channels,'' \emph{Phys. Rev. Lett.}, vol.~77, no.~13, p. 2818, 1996.

\bibitem{brand2020efficient}
S.~Brand, T.~Coopmans, and D.~Elkouss, ``Efficient computation of the waiting time and fidelity in quantum repeater chains,'' \emph{{IEEE} J. Sel. Areas Commun.}, vol.~38, no.~3, pp. 619--639, 2020.

\bibitem{li2021efficient}
B.~Li, T.~Coopmans, and D.~Elkouss, ``Efficient optimization of cutoffs in quantum repeater chains,'' \emph{{IEEE} Trans. Quantum Eng.}, vol.~2, pp. 1--15, 2021.

\bibitem{victora2023entanglement}
M.~Victora, S.~Tserkis, S.~Krastanov, A.~S. de~la Cerda, S.~Willis, and P.~Narang, ``Entanglement purification on quantum networks,'' \emph{Phys. Rev. Res.}, vol.~5, no.~3, p. 033171, 2023.

\bibitem{zang2025entanglement}
A.~Zang, X.-A. Chen, E.~Chitambar, M.~Suchara, and T.~Zhong, ``Entanglement purification in quantum networks: Guaranteed improvement and optimal time,'' arXiv:2505.02286 [quant-ph], 2025.

\bibitem{vasan2024control}
V.~Vasan, A.~Agrawal, A.~Nico-Katz, J.~Horgan, B.~A. Bash, D.~C. Kilper, and M.~Ruffini, ``Control protocol for entangled pair verification in quantum optical networks,'' \emph{arXiv preprint arXiv:2411.07410}, 2024.

\bibitem{ookla2024speedtest}
{Ookla}, ``Speedtest by ookla global fixed network performance—2024-01-01 fixed tiles data,'' 2024, accessed Nov. 4, 2024. [Online]. Available: https://registry.opendata.aws/speedtest-global-performance.

\bibitem{wengerowsky2018entanglement}
S.~Wengerowsky, S.~K. Joshi, F.~Steinlechner, H.~H\"{u}bel, and R.~Ursin, ``An entanglement-based wavelength-multiplexed quantum communication network,'' \emph{Nature}, vol. 564, no. 7735, pp. 225--228, 2018.

\bibitem{jacinto2025network}
H.~Jacinto, {\'E}.~Gouzien, and N.~Sangouard, ``Network requirements for distributed quantum computation,'' \emph{arXiv preprint arXiv:2504.08891}, 2025.

\bibitem{kreuter2005experimental}
A.~Kreuter, C.~Becher, G.~P.~T. Lancaster, A.~B. Mundt, C.~Russo, H.~H\"affner \emph{et~al.}, ``Experimental and theoretical study of the 3d d\textsubscript{2}-level lifetimes of \textsuperscript{40}ca\textsuperscript{+},'' \emph{Phys. Rev. A}, vol.~71, no.~3, p. 032504, 2005.

\bibitem{wang2021single}
P.~Wang, C.-Y. Luan, M.~Qiao, M.~Um, J.~Zhang \emph{et~al.}, ``Single ion qubit with estimated coherence time exceeding one hour,'' \emph{Nat. Commun.}, vol.~12, no.~1, p. 233, 2021.

\bibitem{rancic2018coherence}
M.~Ran\v{c}i\'{c}, M.~P. Hedges, R.~L. Ahlefeldt, and M.~J. Sellars, ``Coherence time of over a second in a telecom-compatible quantum memory storage material,'' \emph{Nat. Phys.}, vol.~14, no.~1, pp. 50--54, 2018.

\bibitem{maurer2012room}
P.~C. Maurer, G.~Kucsko, C.~Latta, L.~Jiang, N.~Y. Yao \emph{et~al.}, ``Room-temperature quantum bit memory exceeding one second,'' \emph{Science}, vol. 336, no. 6086, pp. 1283--1286, 2012.

\bibitem{milul2023superconducting}
O.~Milul, B.~Guttel, U.~Goldblatt, S.~Hazanov, L.~M. Joshi \emph{et~al.}, ``Superconducting cavity qubit with tens of milliseconds single-photon coherence time,'' \emph{PRX Quantum}, vol.~4, no.~3, p. 030336, 2023.

\bibitem{reagor2016quantum}
M.~Reagor, W.~Pfaff, C.~Axline, R.~W. Heeres, N.~Ofek \emph{et~al.}, ``Quantum memory with millisecond coherence in circuit qed,'' \emph{Phys. Rev. B}, vol.~94, no.~1, p. 014506, 2016.

\bibitem{singh2020using}
H.~Singh, Arvind, and K.~Dorai, ``Using a lindbladian approach to model decoherence in two coupled nuclear spins via correlated phase damping and amplitude damping noise channels,'' \emph{Pramana – J. Phys.}, vol.~94, pp. 1--10, 2020.

\bibitem{breuer2002theory}
H.-P. Breuer and F.~Petruccione, \emph{The Theory of Open Quantum Systems}.\hskip 1em plus 0.5em minus 0.4em\relax Oxford, U.K.: Oxford Univ. Press, 2002.

\bibitem{barchielli2009quantum}
A.~Barchielli and M.~Gregoratti, \emph{Quantum Trajectories and Measurements in Continuous Time: The Diffusive Case}.\hskip 1em plus 0.5em minus 0.4em\relax Berlin, Germany: Springer, 2009, vol. 782.

\bibitem{bennett1996mixed}
C.~H. Bennett, D.~P. DiVincenzo, J.~A. Smolin, and W.~K. Wootters, ``Mixed-state entanglement and quantum error correction,'' \emph{Phys. Rev. A}, vol.~54, no.~5, p. 3824, 1996.

\bibitem{van2008system}
R.~Van~Meter, T.~D. Ladd, W.~J. Munro, and K.~Nemoto, ``System design for a long-line quantum repeater,'' \emph{{IEEE/ACM} Trans. Netw.}, vol.~17, no.~3, pp. 1002--1013, 2008.

\bibitem{bali2025routing}
R.~Bali, A.~N. Tittelbaugh, S.~L. Jenkins, A.~Agrawal, J.~Horgan, M.~Ruffini, D.~C. Kilper, and B.~A. Bash, ``Routing and spectrum allocation in broadband quantum entanglement distribution,'' \emph{{IEEE} J. Sel. Areas Commun.}, 2025.

\bibitem{sansa2022visible}
A.~Sansa~Perna, E.~Ortega, M.~Gr\"{a}fe, and F.~Steinlechner, ``Visible-wavelength polarization-entangled photon source for quantum communication and imaging,'' \emph{Appl. Phys. Lett.}, vol. 120, no.~7, p. 071102, 2022.

\bibitem{zwerger2014robustness}
M.~Zwerger, H.~Briegel, and W.~D{\"u}r, ``Robustness of hashing protocols for entanglement purification,'' \emph{Phys. Rev. A}, vol.~90, no.~1, p. 012314, 2014.

\end{thebibliography}

\end{document}